\documentclass{emulateapj}

\usepackage{amsmath}
\usepackage{graphicx}
\usepackage{rotating}
\usepackage{multirow}

\def\G{\Gamma}
\def\g{\gamma}

\def\eps{\epsilon}

\newcommand{\beq}{\begin{equation}}
\newcommand{\eeq}{\end{equation}}
\newcommand{\bea}{\begin{eqnarray}}
\newcommand{\eea}{\end{eqnarray}}

\slugcomment{to appear in ApJ}

\shorttitle{Coasting-in-Wind Model for GRB X-Ray Plateaus}
\shortauthors{Shen \& Matzner}

\begin{document}

\title{Coasting external shock in wind medium: an origin for the X-ray plateau decay component in {\it Swift} GRB afterglows}

\author{Rongfeng Shen and Christopher D. Matzner}
\email{rfshen@astro.utoronto.ca}
\email{matzner@astro.utoronto.ca}
\affil{Department of Astronomy \& Astrophysics, University of Toronto, M5S 3H4, Canada}

\begin{abstract}
The plateaus observed in about one half of the early X-ray afterglows are the most puzzling feature in gamma-ray bursts (GRBs) detected by {\it Swift}. By analyzing the temporal and spectral indices of a large X-ray plateau sample, we find that 55\% can be explained by external, forward shock synchrotron emission produced by a relativistic ejecta coasting in a $\rho \propto r^{-2}$, wind-like medium; no energy injection into the shock is needed. After the ejecta collects enough medium and transitions to the adiabatic, decelerating blastwave phase, it produces the post-plateau decay. For those bursts consistent with this model, we find an upper-limit for the initial Lorentz factor of the ejecta, $\Gamma_0 \leq 46 ~(\eps_e/0.1)^{-0.24} (\eps_B/0.01)^{0.17}$; the isotropic equivalent total ejecta energy is $E_{\rm iso}\sim 10^{53} ~(\eps_e/0.1)^{-1.3} (\eps_B/0.01)^{-0.09} (t_b/10^4 \, {\rm s})$ erg, where $\eps_e$ and $\eps_B$ are the fractions of the total energy at the shock downstream that are carried by electrons and the magnetic field, respectively, and $t_b$ is the end of the plateau. Our finding supports Wolf--Rayet stars as the progenitor stars of some GRBs. It raises intriguing questions about the origin of an intermediate-$\Gamma_0$ ejecta, which we speculate is connected to the GRB jet emergence from its host star. For the remaining 45\% of the sample, the post-plateau decline is too rapid to be explained in the coasting-in-wind model, and energy injection appears to be required.
\end{abstract}

%\date{}


\maketitle

%%%%%%%%%%%%%%%%%%%%%%%%%%%%%%%%%%%%%%%%%%%%%%%%%%%%%%%%%%%%%%%%%%%%%%%%%%%%
\section{Introduction}

Among the new features in early X-ray afterglows recently discovered by {\it Swift}, the X-ray plateau is the most troublesome to explain; see Zhang 2007 for a review. There is a long list of proposed explanations, including a prior emission model (Ioka et al. 2006; Yamazaki 2009), an evolving microphysical parameter model (Panaitescu et al. 2006; Ioka et al. 2006), a two-component (Granot et al. 2006) or multi-component (Toma et al. 2006) jet model, a slow energy transfer from the magnetized ejecta to the ambient medium (Kobayashi \& Zhang 2007), a reverse-shock dominated afterglow model (Uhm \& Beloborodov 2007; Genet et al. 2007), a circumburst-dust-scattered echo of prompt X-rays (Shao \& Dai 2007; but see Shen et al. 2009), and a scattered internal shock (Shen, Barniol Duran \& Kumar 2008) or external shock emission (Panaitescu 2008) model. Most fail to explain all the properties of the X-ray plateau. The leading model invokes energy injection to refresh the external shock.  In this scenario the injected energy comes either from a long-lasting central engine activity (e.g., Dai \& Lu 1998a, b; Zhang \& M\'{e}sz\'{a}ros 2001; Dai 2004; Yu \& Dai 2007) or from a slower portion of an outflow with a broad initial Lorentz factor (LF) distribution (e.g., Granot \& Kumar 2006). However, the standard refreshed shock model faces several serious issues. First, the flat slope ($\sim t^{-0.5}$) and late ending time ($\sim 10^4$ s) of the plateau feature imply that the total energy carried by the later or the slower ejecta is much larger than that of the prompt, fast component which gives rise to $\g$-rays. Second, the abrupt end to the plateau phase poses a serious theoretical challenge to models for the origin of the injected energy. The late-activity scenario also suffers from the fact that prompt $\g$-rays, which are comparable in energy to the late-time injection, emerge so much earlier.  The implied radiative efficiency can exceed 90\% (e.g., Ioka et al. 2006; Zhang et al. 2007), much higher than the $\sim 1\%$ efficiencies attained by the popular internal-shock model. We note that if the energy injection is made by an e$^{\pm}$ pair dominated outflow, such as a dissipated wind from a millisecond magnetar (Dai 2004; Yu \& Dai 2007; Yu, Liu \& Dai 2007; Mao et al. 2010), the wind-decelerating relativistic reverse shock radiates more efficiently than in the case of a baryon-dominated injecting outflow and might even dominate over the forward shock emission; this alleviates the energy budget and the required efficiency. In that model, the X-ray plateau ending time corresponds to the spin-down time of the magnetar.

Some afterglow features under the fireball shock model in a $\rho \propto r^{-k}$ medium, where $k$ is a constant, have been described earlier by Dai \& Lu (1998c) and M\'{e}sz\'{a}ros, Rees \& Wijers (1998). Chevalier \& Li (1999, 2000) have studied specific behavior of afterglow due to the blastwave expanding into the wind of a Wolf--Rayet star, corresponding to $k=2$. Panaitescu \& Kumar (2000) studied the afterglow light curves both in a constant density ($k=0$) interstellar medium (ISM) and in a wind medium. Models of late-time ($t \gtrsim 10^4$ s) afterglow data (e.g., Panaitescu \& Kumar 2002; Schulze et al. 2011; Oates et al. 2011) have inferred that afterglows consistent with ISM are more numerous than those consistent with a wind medium.
%\footnote{\bf The main reason that modeling of the late-time data found more cases of $k=0$ can be seen in Figure \ref{fig:ax2-bx2}. For the spectral regime $\max(\nu_m, \nu_c) < \nu_X$, the X-ray data is degenerate of the medium type. For the other spectral regime $\nu_m < \nu_x < \nu_c$, more data are consistent with $k=0$ than with $k=2$. Moreover, in the late-time optical band, in average the data have slightly harder spectral index than in X-ray band while the temporal decay index doesn't show systematic difference between the two bands (see Table B.1 and B.2 in Schulze et al. (2011). Thus, the model of $k=0$ and $\nu_m < \nu_o < \nu_c$ is the most consistent with the optical data. From the two facts combined together, one can easily conclude that the model that is most consistent with the two-band data is: $k=0$ with $\nu_m < \nu_o < \nu_x < \nu_c$ or $\nu_m < \nu_o < \nu_c < \nu_x$.}

In the external shock model for gamma-ray burst (GRB) afterglows, relativistic ejecta coasts freely before it collects $1/\G_j$ times its rest mass from the circum-burst medium and decelerates, where $\G_j$ is its initial LF. In the past, afterglow studies have concentrated mainly on the decelerating phase, in which the flatness of the X-ray plateau decay slope is difficult to explain. In this paper we look at the early, coasting phase. If the medium density is constant with radius, the light curve in this phase should rise as $t^{2-3}$, simply because the total number of emitting particles increases steeply with time. A feature consistent with such a rise was only detected for a few cases (e.g., Molinari et al. 2007). Such a rise may often be difficult to detect if it finishes so early that it is buried below the prompt emission tail, which is dimming but still brighter than the rising afterglow.

In a wind medium, however, the external shock light curve in the coasting phase is flat or slowly decaying, if the observing frequency is above both the synchrotron injection frequency and the cooling frequency (see \S \ref{sec:model} and Table \ref{tab:beta-alpha}). Waxman (2004) discussed a mildly relativistic shock coasting in the wind for explaining the X-ray plateau in GRB 980425 / SN 1998bw (Kouveliotou et al. 2004) long before the {\it Swift} era. In this paper we investigate systematically the scenario in which the {\it Swift}-observed X-ray plateau is produced in the coasting phase of the ejecta plunging into a wind medium, and the `normal' decay following the plateau corresponds to the subsequent decelerating phase of the ejecta in the same wind medium; we refer to this as the ``coasting-in-wind'' model. We assume the synchrotron emission of the external forward shock is the dominating emission component.

We start with reviewing in \S \ref{sec:model} the dynamics and radiation properties of the external forward shock in a general medium, for both the coasting and decelerating phases of the ejecta. Then in \S \ref{sec:Sample} we model a large sample of X-ray plateau data with the coasting-in-wind model, and in \S \ref{sec:Constraints} we derive constraints on the initial LF and the isotropic equivalent energy of the ejecta. Summary of the results and discussion of the implications are given in \S \ref{sec:Discussion}.

%%%%%%%%%%%%%%%%%%%%%%%%%%%%%%%%%%%%%%%%%%%%%%%%%%%%%%%%%%%%%%%%%%%%%%%%%%%%%%%%%
\section{Emission from external forward shock in a generalized medium} \label{sec:model}

In this work we use the simplest self-consistent assumptions and approximations, including assumptions that all electrons are accelerated into a single power-law distribution, and that the magnetic and electron energy densities are fixed fractions of the post-shock energy density. Simulations of relativistic shock acceleration (Spitkovsky 2008; Sironi \& Spitkovsky 2009, 2011) show that only a small fraction ($\sim 1\%$) of downstream electrons are accelerated in a power-law distribution carrying $\sim 10\%$ of downstream energy. This can be modeled using a fixed  $\eps_e$ (defined below), unless in reality either the number fraction or the energy fraction of non-thermal electrons changes strongly with the shock LF (e.g., see \S\ref{sec:Optical}). The standard afterglow modeling also assumes that electron Larmor radii are smaller than the coherence length of the magnetic field so that synchrotron radiation applies; otherwise the emission is in the so-called `jitter' regime (Medvedev 2000).

After crossing the shock, electrons are accelerated into a power-law energy distribution with an index $p$. Those non-thermal electrons carry a small fraction, $\eps_e$, of the internal energy  downstream. The minimum electron energy in this power law is
\beq
\g_m = \begin{cases}
   f_p \eps_e \G \mu_e m_p/m_e, & {\rm for} ~~ 2 < p, \\
   \left(f_p \eps_e \G  \g_M^{p-2} \mu_e m_p/m_e \right)^{1/(p-1)}, & {\rm for} ~~ 1 < p < 2,
  \end{cases}
\eeq
where $\mu_e$ is the number of nucleons per electron, $f_p=|p-2|/(p-1)$ derives from integrating the number density over the power law (e.g., Sari et al. 1998), and for the second case $\g_M=\sqrt{3 e/(\sigma_T B)}$ is the maximum electron energy in the power law (Dai \& Cheng 2001). Here $\sigma_T$ is the Thomson cross section, $e$ is the electron charge and $B$ is the magnetic field, whose  strength $B=\sqrt{32\pi \eps_B \G^2 \rho c^2}$ is determined by the fraction, $\eps_B$, of the downstream energy which goes to the magnetic field. The accelerated electrons emit synchrotron radiation. Another critical electron energy is the cooling energy above which electrons lose significant energy in the dynamical time:
\beq
\g_c = \frac{6\pi m_e c (1+z)} {(1+Y) \sigma_T \G B^2 t},
\eeq
where $z$ is the GRB host red shift and $Y$ is the synchrotron self Compton (SSC) parameter; we neglect the SSC loss, i.e., $Y= 0$. Above $\g_c$, cooling modifies the power law slope of electrons distribution. Radiative losses are necessarily important when $p<2$, because in this case the electron energy is concentrated at very high energies above $\g_c$; conversely when $p>2$ synchrotron losses are important only when $\g_c < \g_m$.

For the dynamical model, we consider a relativistic ejecta of isotropic equivalent energy $E_{\rm iso}$ and LF $\G_j$ plunging into the circum-burst medium. The medium density profile is generally assumed to be a power law $\rho = A r^{-k}$, where for the steady stellar wind case, $A_{k=2} = \dot{M}_w/(4\pi v_w)= 5\times10^{11} A_*$ g cm$^{-1}$  (Chevalier \& Li 1999); the fiducial value $A_*=1$ would arise for a wind of mass loss rate $\dot{M}_w= 10^{-5} M_{\odot}$ yr$^{-1}$ and speed $v_w= 10^3$ km s$^{-1}$.  In the case of a uniform density (`ISM') medium, $A_{k=0}=\rho$.

A pair of shocks is produced, with the forward shock (LF $\G_{\rm sh}$) moving into the medium and the reverse shock into the ejecta. The shocked fluid region comprises the forward-shocked medium and reverse-shocked ejecta, which have approximately the same LF $\G(t)\approx \G_{\rm sh}/\sqrt{2}$ but are separated by a contact discontinuity.  $\G(t)$ is given by pressure balance at the discontinuity as $\G= \G_j/(1+2\G_j/a^{1/2})^{1/2}$, where $a= \rho_j'/\rho(r)$ is the rest-frame density ratio between the ejecta and the ambient medium (e.g., Panaitescu \& Kumar 2004b). In the limit of $a \gg \G_j^2$, $\G \simeq \G_j$. It can be shown that, before the deceleration, i.e., the reverse shock passage, of the ejecta, the dense shell condition $a \gg \G_j^2$ is satisfied.\footnote{The ambient density is $\rho(r) = (3-k)M(r)/(4\pi r^3)$, and the rest-frame jet density is $\rho_j'= M_j/(4\pi r^2 \Delta_j')$, where $M_j$ is the isotropic ejecta mass and  $\Delta_j'$ is the ejecta width in the rest frame.   For a late deceleration time, which is appropriate for our coasting-in-wind model, the ejecta width is determined by radial spreading, so $\Delta_j' \sim r/\G_j$.  Thus $a =\rho_j'/\rho(r)  \sim \G_j M_j/M(r)$.  Prior to deceleration, $M_j/\G_j \gg M(r)$ (cf. Equation \ref{eq:Eiso}), so $a \gg \G_j^2 $; however $a \sim \G_j^2$ at deceleration.}
%\footnote{Before the deceleration, the swept-up medium mass is negligible compared with the initial ejecta rest mass divided by $\G_j$: $M_m= 4\pi A r^{3-k}/(3-k) \ll M_j/\G_j$. The ejecta rest frame density $\rho_j'= M_j/(4\pi r^2 \Delta_j')$, where $\Delta_j'$ is the ejecta rest frame width. For a late deceleration time, which is appropriate for our coasting-in-wind model, the ejecta width is determined by the radial spreading, i.e., $\Delta_j' \sim r/\G_j$. Thus, we have $a= \rho_j'/\rho(r) \sim \G_j M_j /(4\pi A r^{3-k}) \gg \G_j^2$.}
In the coasting-in-wind model $\G_j$ takes a single value rather than a distribution, so that $\G(t)$ has a definite initial value $\G_0 \simeq \G_j$.

Right after the reverse shock passage, about half of the initial ejecta total energy is transferred to the shocked medium and ejecta. The combined shocked region enters the Blandford \& McKee (1976) self-similar solution, in which $\G(t)$ declines.  For convenience, we denote phase 1 as the period in which the reverse shock is crossing the ejecta, and phase 2 the later period of blastwave deceleration. In phase 1, since usually the reverse shock is too weak to account for the X-ray emission, we consider the forward-shocked region only whenever radiation properties are concerned.

In phase 2 the shock compresses the density and enhances the energy per unit mass both by a factor of $\sim \G$; energy conservation requires (Cohen et al. 1998; also see Wu et al. 2005 for similar treatment of radiative loss)
\beq    \label{eq:Eiso}
E_{\rm iso} = \frac{8\pi A \G_{\rm sh}^2 r^{3-k} c^2} {17-4k+\epsilon}\left(1-\frac\epsilon3\right)
\eeq
where, to account for synchrotron losses,
\beq \label{eq:epsilon_approx}
\epsilon \simeq \left\{  \begin{array}{cl}
0, & {\rm if~} p>2 {\rm ~~and~} \g_m < \g_c  \\
\frac12 (\epsilon_e + \epsilon_e^{3}), & {\rm otherwise}
\end{array}
\right.
\eeq
is the fraction of post-shock energy, if any, radiated away; approximation (\ref{eq:epsilon_approx}) matches the more complicated exact solution by Cohen et al (1998) to within 5\%. If $p>2$ and $\g_m < \g_c$ so that $E_{\rm iso}$ is conserved in phase 2, $\Gamma\propto r^{-(3-k)/2}\propto t^{-(3-k)/(8-2k)}$; otherwise $\Gamma^2 \propto r^{-m}$ where (Cohen et al. 1998)
\beq \label{eq:m_in_Gamma^2}
m = \frac{(1+\epsilon)^2+3(1+\epsilon)(4-k)-4} {3-\epsilon}
\eeq
which in the adiabatic limit becomes $m_{\epsilon=0}= 3-k$. In any phase of continuous motion with $\G^2\propto r^{-m}$, the observer's time $t$ is given by
\beq \label{eq:t}
t = \frac{1+z} {1+m} \frac{r} {c_t  \G^2 c}  ~\longrightarrow ~ \frac{1+z}{4-k} \frac{r}{c_t \G^2 c},
\eeq
where the arrow evaluates $m\rightarrow (3-k)$ for a decelerating, adiabatic blastwave.
For adiabatic blastwaves, we find  $\Gamma \propto r^{-1/2} \propto t^{-1/4}$ in a wind medium where $t =(1+z) r/(2 c_t \Gamma^2 c)$, and $\Gamma \propto r^{-3/2} \propto t^{-3/8} $ in a uniform medium where $t =(1+z) r/(4 c_t \Gamma^2 c)$. The prefactor $c_t= 4$ if most of the observed emission is from a thin layer of shocked fluid right behind the shock front and on the line-of-sight axis (Sari 1997; Dai \& Lu 1998c; Chevalier \& Li 2000). Emission from off axis and from the fluid further downstream arrives somewhat later, so the effective $c_t$ is somewhat smaller (Waxman 1997; Sari 1998; Panaitescu \& Mesazaros 1998). We adopt $c_t= 2$.
%\footnote{Generally, one can write the relation of observer's time, radius and LF as $t= (1+z)r/[(4-k) c_t \Gamma^2 c]$. In Eq. (\ref{eq:t}) the choice of the numerical coefficient $c_t= 4$ is based on the assumption that most emission is from a thin layer of shocked fluid right behind the shock front at the line of sight (Sari 1997; Dai \& Lu 1998c). Waxman (1997) argues that the exact value of $c_t$ is dependent on the radial distribution of the electron energy and magnetic energy in the shocked region further away from the shock front, and also depends on the emission spectrum since the observer also sees the photons from high altitudes of the equal arrival time surface. Considering these effects will make $c_t$ smaller. Sari (1998) considered the equal arrival time surface effect in the case of uniform density ($k=0$) and found $c_t \sim 1$. Considering the same effect, however, Panaitescu \& Mesazaros (1998) found $c_t$ is just close to 4. We note that Panaitescu \& Kumar (2000) use $c_t= 2$ even for the line-of-sight photons.}

The observed synchrotron frequency for electrons with energy $\g$, averaged over an isotropic distribution of the electron's pitch angle, is $\nu= 3 \G \g^2 e B / [16 m_e c (1+z)]$. The critical frequencies corresponding to $\g_m$ and $\g_c$ are $\nu_m$ and $\nu_c$, respectively. The peak synchrotron specific power for a single electron is $P_{\nu, \max} = 8 m_e c^2 \sigma_T \G B / (9 \pi e)$, independent of the electron energy. The peak flux density in the observed $F_{\nu}$ spectrum is $F_{\nu,\max}= N P_{\nu, \max} (1+z) /(4\pi d_L^2)$, where $N = 4\pi A r^{3-k} / [(3-k)m_p]$ is the total number of electrons in the swept-up medium, and $d_L$ is the luminosity distance.

Given the above formulae, we can find the following time evolution laws for the case of $p> 2$:
\beq    \label{eq:scaling-pha-1}
\nu_m \propto t^{-k/2}, ~~ \nu_c \propto t^{3k/2-2}, ~~ F_{\nu,\max} \propto t^{3-3k/2}
\eeq
for phase 1, and
\bea    \label{eq:scaling-pha-2}
\nu_m \propto t^{-\frac{k+4m}{2(m+1)}}, ~ \nu_c \propto t^{\frac{3k-4}{2(m+1)}}, ~ F_{\nu,\max} \propto t^{-\frac{3k-6+2m}{2(m+1)}}
\eea
for phase 2. For $1<p<2$, the only difference is on $\nu_m$: $\nu_m \propto t^{\frac{-k}{2(p-1)}}$ in phase 1, and $\nu_m \propto t^{-\frac{m(p+2)+k}{2(m+1)(p-1)}}$ in phase 2.
Note that for wind medium, $\nu_m$ decreases and $\nu_c$ increases monotonically with time in both phases.

Following from the above evolution laws the observed flux density $F_{\nu}$ evolves according to $F_{\nu}(t) \propto \nu^{-\beta} t^{-\alpha}$ (e.g., Sari, Piran \& Narayan 1998). The spectral slope $\beta$ is determined by $p$ and the relative locations of $\nu$, $\nu_m$ and $\nu_c$. Therefore, a specific numerical relation between $\alpha$ and $\beta$, the so-called ``closure relation'', exists depending on which type of medium and which dynamical phase the external shock is in. We refer the reader to Zhang \& M\'{e}sz\'{a}ros (2004) for a summary of these relations in the decelerating phase for both uniform and wind media. For demonstration purpose, we list in Table \ref{tab:beta-alpha} the closure relations for both the coasting phase and the decelerating phase in a $k=2$ wind medium.

%%%%%%%%%%%%%%%%%%%%%%%%%%%%%%%%Begin Table 1%%%%%%%%%%%%%%%%%%%%%%%%%%%%%%%%%%%%%%%%%%%%
\begin{table*}
\caption{The spectral and temporal indices of the forward shock synchrotron emission ($F_{\nu} \propto \nu^{-\beta} t^{-\alpha}$) in the coasting-in-wind model for the wind density index $k=2$. }      \label{tab:beta-alpha}
\vspace{10pt}
\begin{center}
\begin{tabular}{lccccc} \hline \hline
\multirow{2}{*}{Spectral regime} & \multirow{2}{*}{$\beta$} & \multicolumn{2}{c}{$\alpha$ ($p>2$)} & \multicolumn{2}{c}{$\alpha$ ($p<2$)}  \\
 & & phase 1 & phase 2 & phase 1 & phase 2 \\ \hline

Slow cooling ($\nu_m < \nu_c$) & & & & & \\
$\nu < \nu_m$ & $-\frac{1}{3}$ & $-\frac{1}{3}$ & 0 & $-\frac{1}{3(p-1)}$ & $-\frac{5(2-p)}{12(p+1)}$ \\
$\nu_m < \nu < \nu_c$ & $\frac{p-1}{2}$ & $\frac{p-1}{2}$ & $\frac{3p-1}{4}$ & $\frac{1}{2}$ & $\frac{p+8}{8}$ \\
\hline
Fast cooling ($\nu_c < \nu_m$) & & & & & \\
$\nu < \nu_c$ & $-\frac{1}{3}$ & $\frac{1}{3}$ & $\frac{2}{3}$ & $\frac{1}{3}$ & $\frac{2}{3}$ \\
$\nu_c < \nu < \nu_m$ & $\frac{1}{2}$ & $-\frac{1}{2}$ & $\frac{1}{4}$ & $-\frac{1}{2}$ & $\frac{1}{4}$ \\
\hline
$\max(\nu_m, \nu_c) < \nu$ & $\frac{p}{2}$ & $\frac{p-2}{2}$ & $\frac{3p-2}{4}$ & 0 & $\frac{p+6}{8}$ \\ \hline
\hline
\end{tabular}
\end{center}
\tablecomments{Phases 1 and 2 are the coasting and decelerating phases of the blast wave, respectively.}
\end{table*}
%%%%%%%%%%%%%%%%%%%%%%%%%%%%%%%%End TAble 1%%%%%%%%%%%%%%%%%%%%%%%%%%%%%%%%%%%%%%%%%%%%%%

%%%%%%%%%%%%%%%%%%%%%%%%%%%%%%%%%%%%%%%%%%%%%%%%%%%%%%%%%%%%%%%%%%%%%%%%%%%%%%%%%%%%%%%%%%%%%%
\begin{figure}
\centerline{
\includegraphics[width=10cm, angle=0]{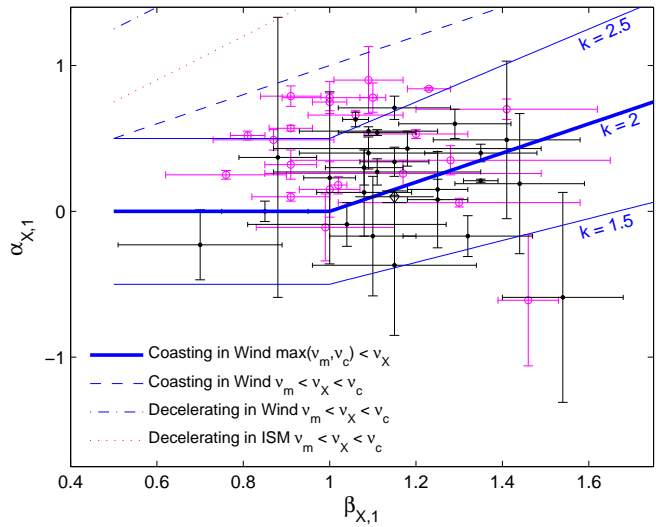}
}
\caption{X-ray decay index vs. spectral index during the plateau phase. The lines are the closure relations predicted by various models. The prediction from the coasting-in-wind model with $\max(\nu_m, \nu_c) < \nu$ (\textit{thick solid} line) is the most capable one that accounts the data. The predictions from the same model with slight variations in $k$ (\textit{thin solid} lines) are also plotted. The data in \textit{open} circles (magenta color) are those whose post-plateau decays are inconsistent with the decelerating phase (phase 2) of the coasting-in-wind model (see Figure \ref{fig:ax2-bx2} and in the text). GRB 061202 that shows a strong spectral evolution is plotted as a diamond.}  \label{fig:ax1-bx1}
\end{figure}
%%%%%%%%%%%%%%%%%%%%%%%%%%%%%%%%%%%%%%%%%%%%%%%%%%%%%%%%%%%%%%%%%

%%%%%%%%%%%%%%%%%%%%%%%%%%%%%%%%%%%%%%%%%%%%%%%%%%%%%%%%%%%%%%%%%%%%%%%%%%%%%%%%%%%%%%%%%%%%%
\begin{figure}
\centerline{
\includegraphics[width=10cm, angle=0]{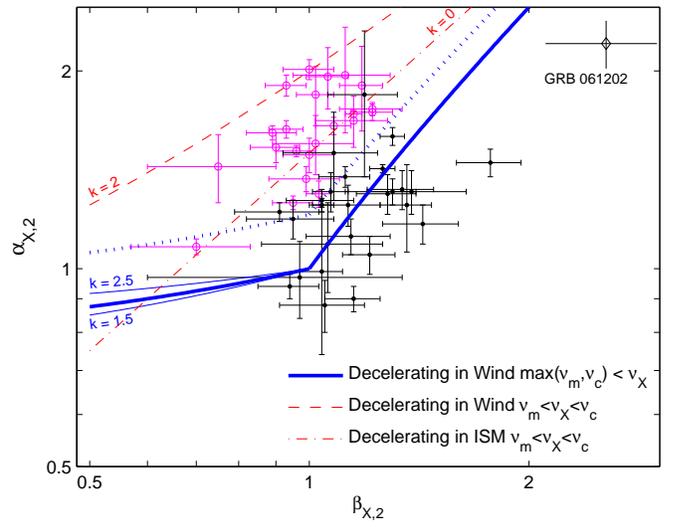}
}
\caption{X-ray temporal index vs. spectral index during the post-plateau phase. The \textit{solid} lines are for the decelerating phase of the coasting-in-wind model. Note that for a decelerating blastwave with $\max(\nu_m, \nu_c) < \nu$ and $p > 2$, $\alpha$ is independent of $k$. The data in \textit{open} circles (magenta color) are for the inconsistent sub-sample. The \textit{thick dotted} line (blue color) is the prediction for phase 2 of the coasting-in-wind model considering the radiative loss of blastwave energy (requiring $\nu_c < \nu_m$ when $p>2$, and assuming $\eps_e=0.3$); it should be taken as an upper limit of $\alpha_{X,2}$ for a given $\beta_{X,2}$ within this model because of the high $\eps_e$ value assumed. The exceptional burst that shows a spectral evolution is labeled.}	 \label{fig:ax2-bx2}
\end{figure}
%%%%%%%%%%%%%%%%%%%%%%%%%%%%%%%%%%%%%%%%%%%%%%%%%%%%%%%%%%%%%%%%%%%%%%%%%%%%%%%%%%%%%%%%%%

%%%%%%%%%%%%%%%%%%%%%%%%%%%%%%%%%%%%%%%%%%%%%%%%%%%%%%%%%%%%%%%%%%%%%%%%%%%%
\section{Sample and data analysis}      \label{sec:Sample}

We use the data from a sample of 53 X-ray plateaus compiled by Liang et al. (2007). This sample is provided with fitted spectral and temporal slopes of the plateau component, $\beta_1$ and $\alpha_1$, and of the normal decay component following the plateau, $\beta_2$ and $\alpha_2$, and the time of the transition between the two components $t_b$. We omit four bursts whose phase 2 decay slopes are too steep ($\alpha_2 \approx 3$ for GRB 060413, 060522 and 060607A; $\alpha_2 \approx 9$ for GRB 070110) to be explained by external shocks; for them the best explanation is that some internal dissipation of a delayed central engine ejecta is responsible for both the abrupt step-like decline in phase 2 and the prior plateau (e.g., Troja et al. 2007; Lyons et al. 2010). This leaves us a sample of 49 bursts.

Figure \ref{fig:ax1-bx1} compares the observed values of $\alpha_1$ and $\beta_1$ against the predicted closure relations for different dynamical, medium and spectral models.  It can be seen that the coasting-in-wind model with $\max(\nu_m,\nu_c) < \nu_X$  most successfully accommodates the majority of data.

Within this model the post-plateau decay corresponds to the decelerating phase (phase 2) of the blast wave in a wind medium; the break cannot be due to a change of spectral regime because there is no spectral evolution ($\beta_2=\beta_1$) in all but one burst --- GRB 061202 (e.g., Liang et al. 2007; Shen et al. 2009). The lack of spectral evolution also signals a requirement for any model, that is, whichever spectral regime phase 1 is in, it remains in phase 2.  We therefore expect a specific relation between $\alpha_2$ and $\beta_2$ depending on the relative locations of $\nu_X$, $\nu_m$ and $\nu_c$.

Figure \ref{fig:ax2-bx2} plots the observed $\alpha_2$ vs. $\beta_2$ with predicted relations superimposed. For the decelerating phase in a wind medium, we see that the spectral regime $\max(\nu_m, \nu_c)<\nu_X$ is the most consistent with the data, which also satisfies the requirement of zero spectral evolution. However, as in Figure \ref{fig:ax1-bx1}, the data in Figure \ref{fig:ax2-bx2} show broad scatter around the prediction of this model. While the scatter in the former figure could be attributed to a slight variation of the density index $k$ from burst to burst, the scatter in the latter cannot --- because $\alpha_2$ has no $k$-dependence in the $\max(\nu_m, \nu_c)<\nu_X$ spectral regime. This is a puzzle for all decelerating blast wave models that do not invoke late energy injection.

\begin{figure}
\centerline{
\includegraphics[width=9cm, angle=0]{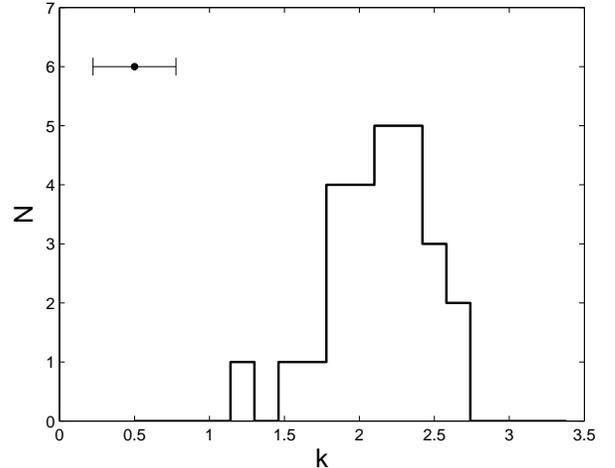}
}
\caption{Distribution of the derived density power-law index $k$ for the sub-sample that are consistent with the coasting-in-wind model in both its phase 1 and phase 2 -- the consistent sub-sample. The top left horizontal bar shows the size of the typical error of $k$.}	\label{fig:k1b}
\end{figure}

In conclusion, we find that the coasting-in-wind model for the plateau, and its subsequent (adiabatic) decelerating stage for the post-plateau decay, can account for 55\% of the sample. We call these bursts the \textit{consistent sub-sample}, and they are plotted as black data points in Figures \ref{fig:ax1-bx1} and \ref{fig:ax2-bx2}. GRB 061202 is consistent with the closure relation predictions of the coasting-in-wind model both during and after the plateau phase, but its strong softening evolution from $\beta_{1}= 1.15\pm0.07$ to $\beta_{2}= 2.55\pm0.44$ cannot be accounted for by a nominal spectral regime change $\nu_m < \nu_X < \nu_c \rightarrow \max(\nu_m, \nu_c) < \nu_X$, nor by any variant of the coasting-in-wind model; we therefore exclude it from the consistent sub-sample. The remaining of the sample, i.e., those located above the coasting-in-wind model prediction by at least 2$\sigma$ in Figure \ref{fig:ax2-bx2}, are marked as open circles (in magenta color) both in Figure \ref{fig:ax2-bx2} and Figure \ref{fig:ax1-bx1}. We call these the \textit{inconsistent sub-sample}. We see in Figure \ref{fig:ax1-bx1} that the inconsistent sub-sample shows no significant difference from the consistent sub-sample in the $\alpha_1$ vs. $\beta_1$ scatter plot. Figure \ref{fig:k1b} shows the distribution of $k$ derived from the consistent sub-sample in phase 1, which is centered around $k=2$.

At face value, this finding seems to contrast with previous studies (e.g., Panaitescu \& Kumar 2002; Schulze et al. 2011;Oates et al. 2011) of late-time ($t \gtrsim 10^4$ s) afterglow data, which concluded that GRBs with $k\simeq 0$ predominate over those with $k\simeq2$. However, the difference should not come as a surprise. Our consistent sub-sample includes only those bursts consistent with the $\max(\nu_m, \nu_c) < \nu_X$ spectral regime  --- as required by our modeling of the plateau data and by the lack of spectral evolution. Since $k$ cannot be determined from the model prediction in phase 2 for this spectral regime (solid line in Figure \ref{fig:ax2-bx2}), we infer $k$ instead from plateau data. Our inferred $k$ values, based as they are on the assumption of a coasting ejecta, are model dependent: we would obtain very different values within a model involving late central engine activity.

As for the inconsistent sub-sample, interpreting these under the $\nu_m < \nu_X < \nu_c$ spectral regime indicates that many are closer to $k=0$ than $k=2$. Two bursts, 050801 and 060714, appear in both the Schulze et al. (2011) sample and our consistent sub-sample; Schulze et al. classify both as $k=0$.   Such overlap is expected, as we describe a burst as `consistent', and derive its $k$ from the coasting-in-wind model, if it is within 2$\sigma$ of the $\max(\nu_m, \nu_c) < \nu_X$ locus in phase 2. Bursts at the upper periphery of this sub-sample could also be interpreted assuming $\nu_m < \nu_X < \nu_c$, and the derived $k$ would be very different.  Note also that our values for the two bursts' post-plateau indices, from Liang et al. (2007), differ slightly from those used by Schulze et al., and the latter authors treat a burst as consistent if it falls within 3$\sigma$ of a model, whereas our threshold is 2$\sigma$.

%\textbf{Previous studies (e.g., Panaitescu \& Kumar 2002; Schulze et al. 2011) of late-time ($t \gtrsim 10^4$ s) afterglow data concluded with more $k=0$ cases than $k=2$ cases. This difference from our conclusion is due to the following reason. Our work has a prior preference for medium type and spectral regime [wind with $\max(\nu_m, \nu_c) < \nu_X$] arising from modeling the plateau data, while others' don't. Thus, we discarded the upper 45\% portion of the post-plateau data (see Figure \ref{fig:ax2-bx2}) into the inconsistent sub-sample. In others' work, most of this portion will be easily classified as ISM cases (dot-dashed line in Figure \ref{fig:ax2-bx2}) and only a small number of bursts will be classified as wind cases (dashed line), and many in our consistent sub-sample would otherwise be classified as indiscernable for medium type because the closure relation there (solid line) is degenerate with $k=0$ and $k=2$. Two bursts, 050801 and 060714, appear in both Schulze et al. (2011)'s sample and our consistent sub-sample, and both bursts are concluded by Schulze et al. as in ISM cases. This difference from our conclusion arises from different fitting data products being used: these two bursts are located in Figure \ref{fig:ax2-bx2} between the dot-dashed line and the solid line; the slightly smaller $\alpha_{X,2}$ and/or larger $\beta_{X,2}$ for these two from Liang et al. (2007) that we use put them slightly closer to the solid line than in Schulze et al. (2011).}

Allowing for the radiative loss of blastwave energy in phase 2 can help to increase the size of the consistent sub-sample, but not by much. For instance, in the most radiative case --- $\eps_e=0.3$, and $\nu_c < \nu_m$ for $p>2$ particularly --- the predicted $\alpha_{X,2}$ in Figure \ref{fig:ax2-bx2} is shifted upward by $\approx 0.2$. That freedom could at most increase the consistent sub-sample size to about 2/3 of the total.

To account for the inconsistent sub-sample, one must invoke the energy refreshment in the decelerating blast wave. That is the topic of the next subsection.

%%%%%%%%%%%%%%%%%%%%%%%%%%%%%%%%%%%%%%%%%%%%%%%%%%%%%%%%%%%%%%%%%%%%%%%%%%%%%%%%%%%%%%%%%%%%%%

%%%%%%%%%%%%%%%%%%%%%%%%%%%%%%%%%%%%%%%%%%%%%%%%%%%%%%%%%%%%%%%%%%%%%%%%%%%%%%%%%%%%%%%%%%%%%
%%%%%%%%%%%%%%%%%%%%%%%%%%%%%%%%%%%%%%%%%%%%%%%%%%%%%%%%%%%%%%%%%%%%%%%%%%%%%%%%%%%%%%%%%%

%%%%%%%%%%%%%%%%%%%%%%%%%%%%%%%%%%%%%%%%%%%%%%%%%%%%%%%%%%%%%%%%%%%%%%%%%%%%%%%%%%%%%%%%%%%%%
\begin{figure}
\centerline{
\includegraphics[width=9cm, angle=0]{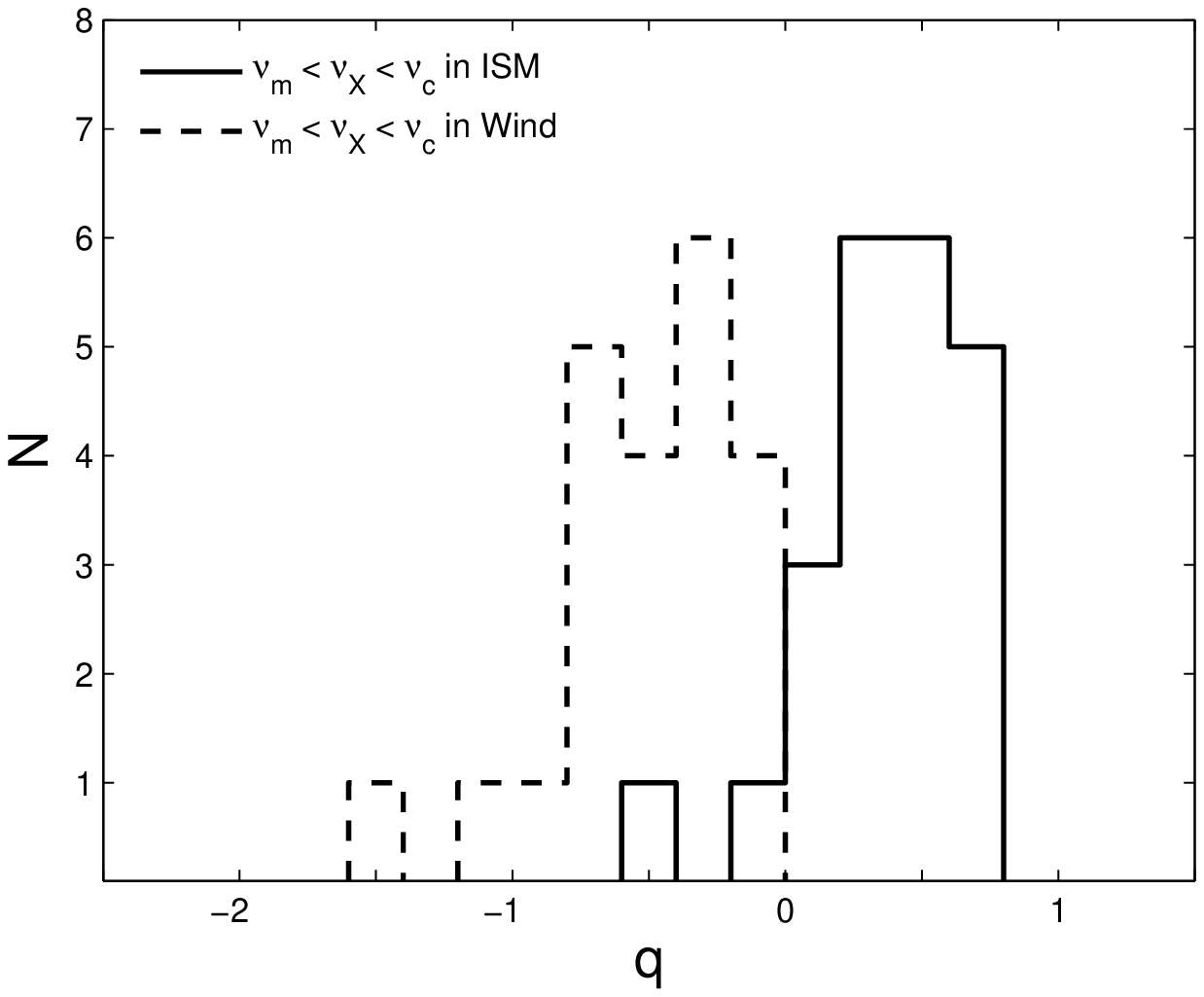}
}
\centerline{
\includegraphics[width=9cm, angle=0]{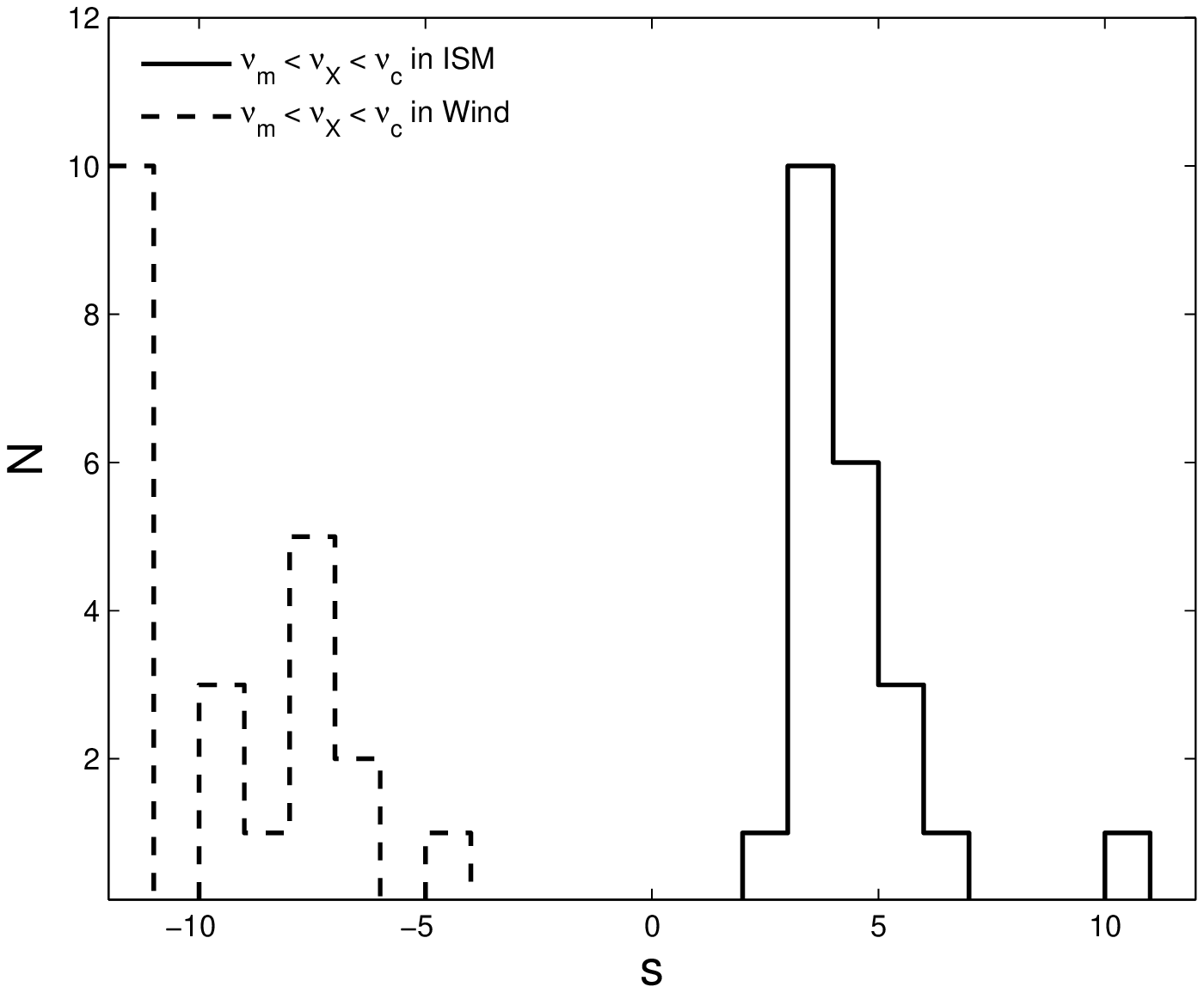}
}
\caption{Distribution of the derived $q$ (\textit{top}) and $s$ (\textit{bottom}) for the inconsistent sub-sample that require a refreshed shock for interpretation, where $q$ is the late central engine luminosity temporal index and $s$ is the catching-up ejecta mass distribution index (see section \ref{sec:refresh}). The spectral regime $\nu_m < \nu_X < \nu_c$ is required by the data in phase 2. For a refreshed shock in a wind medium, the data require unreasonable conditions, i.e., $q < 0$ and $s < 0$, for the two shock refreshment scenarios, respectively. Therefore, ISM is the only allowed medium type when a refreshed shock is considered.}	 \label{fig:q-s}
\end{figure}
%%%%%%%%%%%%%%%%%%%%%%%%%%%%%%%%%%%%%%%%%%%%%%%%%%%%%%%%%%%%%%%%%%%%%%%%%%%%%%%%%%%%%%%%%%

\subsection{Refreshed shock models}     \label{sec:refresh}

There are two primary scenarios in which the blast wave can be refreshed\footnote{A third scenario was suggested by Kobayashi \& Zhang (2007), in which the ejection is brief but is dominated by the Poynting flux. The magnetic energy in the ejecta is not transferred to the shocked ambient medium until the ceasing of the reverse shock. The delayed transfer of the magnetic energy serves as a varied version of the two scenarios mentioned here for the refreshed shock model. However, the process of the magnetic energy transfer in the shocked region is poorly understood. Fan \& Piran (2006) and Mimica et al. (2009) claim that the time scale for the transfer is at most several times the light crossing time of the ejecta, too short to account for the plateau feature.}.

1. The central engine activity is long lasting (Zhang \& M\'{e}sz\'{a}ros 2001) with an outflow luminosity history $L(t) \propto t^{-q}$. Since the LF of the outflow is always much larger than that of the blast wave, this is also the rate at which the total energy of the blast wave gets replenished, i.e., $E_{\rm iso}(t) \propto t^{1-q}$. In order for the late energy injection to be important, $q<1$.

2. The central engine is briefly active, but after the prompt emission phase finishes, the ejecta has a wide range of LF $\G_j$ such that outer shells have larger $\G_j$ and moreover the ejecta mass (or even energy) is dominated by the inner, slower shells, i.e., the mass of ejecta is a power law in $\G_j$ (Rees \& M\'{e}sz\'{a}ros 1998; Panaitescu, M\'{e}sz\'{a}ros \& Rees 1998; Sari \& M\'{e}sz\'{a}ros 2000; Ramirez-Ruiz, Merloni \& Rees 2001): $M(>\G_j) \propto \G_j^{-s}$. This extended ejecta distribution could be the outcome of a GRB jet breaking out of the progenitor star surface or the internal shocks during the prompt emission phase. The coasting-in-wind model can be considered the $s\rightarrow \infty$ limit of this class, in the sense that the ejecta have a single $\G_j$.

In both scenarios the blast wave energy is replenished at some certain rate, so as to cause a flattened plateau temporal slope, and $t_b$ corresponds to the ending of energy injection; the post-plateau decay is the standard decelerating blast wave phase. For any given $\alpha_1$ and a medium type, there is a one-to-one correspondence relation between values of $q$ and $s$ for which the forward shock evolution is identical (although the reverse shock is not), so it is not possible to distinguish between the two scenarios on the basis of X-ray data alone. We find that the inconsistent sub-sample can be accommodated within these models in either a wind or a uniform medium, so long as the observed band is below the cooling break ($\nu_m < \nu < \nu_c$). The lack of spectral evolution implies that this is true in the plateau phase as well. With energy injection a new closure relation exists in phase 1 among $\alpha_1$, $\beta_1$ and $q$ (or $s$), within a given ambient medium (Zhang et al. 2006). From this we can infer the required $q$ (or $s$) value for each burst in the inconsistent sub-sample. The resultant distributions of $q$ and $s$ are shown in Figure \ref{fig:q-s}.

Figure \ref{fig:q-s} shows that within a wind medium the required parameters are unreasonable, in the sense that  $q < 0$ or $s < 0$ for the two varieties of model.  The inconsistent sub-sample can therefore be best accommodated within a refreshed-shock model (of either variety) operating in a uniform medium, and the cooling break must be above the observed band.
Note, however, that in the post-plateau phase a small portion of these bursts appears to decelerate within a wind rather than a uniform medium (see Figure \ref{fig:ax2-bx2}, the dashed line); further examination of these is warranted.
% that this conclusion about the medium type is in group sense; it is possible that a small number of busts might have negative $q$ or $s$ values but not very far from zero and are consistent with a wind medium. This arises from the fact that a few bursts in post-plateau phase with $\nu_m < \nu_X < \nu_c$ are more consistent with wind than with ISM (see Figure \ref{fig:ax2-bx2}). One has to work with bursts individually to pick these bursts out.
The inferred $q$-distribution peaks at $q \sim 0.5$ and the $s$-distribution peaks at $s \sim 3$. This means that in these two refreshed shock scenarios, respectively, the outflow total kinetic energy is dominated either by the late ejecta or by the slower, massive ejecta.

%%%%%%%%%%%%%%%%%%%%%%%%%%%%%%%%%%%%%%%%%%%%%%%%%%%%%%%%%%%%%%%%%%%%%%%%%%%%%%%%%%%%%%%%%%%%%
\subsection{Optical data}       \label{sec:Optical}

A minority (13) of the original X-ray plateau sample in Liang et al. (2007) have simultaneous optical monitoring data; six of them are within our consistent sub-sample. Comparison of the optical light curve with the X-ray light curve shows diversity. Three (GRB 050801, 060714, and 060729) show achromatic breaks around $t_b$, while in the other three the breaks are chromatic. This roughly equal ratio of achromatic to chromatic bursts is about the same for the total 13 X-ray plateaus with optical data in Liang et al. (2007). Among the latter three, GRB 060210 shows an optical break much earlier than $t_b$ (there is no optical coverage on $t_b$ or later), GRB 060526 shows an optical break later than $t_b$, and the optical light curve in GRB 050319 shows no break at all and is consistent with a single power law fit. Can our coasting-in-wind model explain the diverse optical behavior as well?

The achromatic breaks seen in the former three bursts are consistent with the coasting-in-wind model, according to which the break at $t_b$ is of dynamical origin. In these three bursts the optical and X-ray temporal slopes are very similar, both before and after the break (see Figure 2 and Table 2 of Liang et al. 2007). This implies the same physical origin for emission in the two wave bands, which fits well to the coasting-in-wind model. This consistency also applies to the sub-sample compiled by Panaitescu \& Vestrand (2011) whose coupled optical and X-ray light curves both show plateaus (their Figure 4).

An early optical break before $t_b$, such as seen in GRB 060210, can be interpreted as well in the coasting-in-wind model by the passage of $\nu_m$ or $\nu_c$ across the optical band during phase 1; for instance, the spectral regime change $\nu_{o} < \nu_m < \nu_c \rightarrow \nu_m < \nu_{o} < \nu_c$ causes the slope steepen by $\Delta \alpha_o= (3p-1)/6$ (see Table \ref{tab:beta-alpha}; cf. observed $\Delta \alpha_o \approx 1$ in GRB 060210). Note that this change is consistent with the restriction that $\nu_m$ decreases and $\nu_c$ increases in the model. This pure spectral change scenario could also explain those bursts in which the optical shows a peak at the beginning of or during the X-ray plateau (e.g., GRB 060607A and 061121; see Liang et al. 2010 and Panaitescu \& Vestrand 2011). Similarly, a $\nu_m$ or $\nu_c$ passage over $\nu_o$ in phase 2 can explain the optical break later than $t_b$, such as seen in GRB 060526.

The real difficulty lies in explaining the lack of an optical break around $t_b$ as seen in GRB 050319. If the end of X-ray plateau is of a dynamical origin, as postulated in the coasting-in-wind model, it is hard to avoid an achromatic break at $t_b$. Panaitescu et al. (2006) proposed that one solution is to let $\eps_e$ and $\eps_B$ be functions of the blastwave LF $\G$ (also see Panaitescu \& Kumar 2004a):
\beq
\eps_e \propto \G^{-e}, ~~~ \eps_B \propto \G^{-b}.
\eeq
We adopt this treatment here for the coasting-in-wind model, which means in phase 1 $\eps_e$ and $\eps_B$ are still constant, but they are free to evolve with time in phase 2. Note that Panaitescu et al.'s scenario uses energy injection for the X-ray plateau while the coasting-in-wind model does not. In the following, we derive the condition on $e$ and $b$ for the lack of optical break at $t_b$.

We first calculate the X-ray decay slope in phase 2 in the presence of evolving $\eps_e$ and $\eps_B$, since we know its spectral regime is $\max(\nu_m, \nu_c) < \nu_X$:
\beq    \label{eq:ax2eb}
\alpha_{X,2}= \frac{1}{4}+\frac{b}{16}+\frac{(p-1)}{2} \left( \frac{3}{2} -\frac{e}{2} -\frac{b}{8} \right).
\eeq
Next we calculate the optical slope in phase 2. The observed slope $\alpha_o$ in those single power-law optical light curves usually lies in the range of 0.5 -- 0.8. In phase 1, this slope is best consistent with the spectral regime $\nu_m < \nu_o < \nu_c$ for which $\alpha_{o,1}= (p-1)/2$ (see Table \ref{tab:beta-alpha}). This spectral regime should remain in phase 2 since there is no break in the light curve. Therefore, in the presence of evolving $\eps_e$ and $\eps_B$,
\beq
\alpha_{o,2}= \frac{1}{2}-\frac{b}{8}+\frac{(p-1)}{2} \left( \frac{3}{2} -\frac{e}{2} -\frac{b}{8} \right).
\eeq
Then, the equality $\alpha_o= \alpha_{o,1} = \alpha_{o,2}$ poses the following condition for the lack of an optical break:
\beq
\alpha_{o}= \frac{1}{2}-\frac{b}{8}+\alpha_{o} \left( \frac{3}{2} -\frac{e}{2} -\frac{b}{8} \right).
\eeq
In the case of GRB 050319, $\alpha_o \approx 0.5$. The condition becomes
\beq    \label{eq:relation-b-e}
3b+4e= 12.
\eeq
Together with the observed $\alpha_{X,2} \approx 1$ for GRB 050319 and Eq. (\ref{eq:ax2eb}), it gives $e \approx 0$ and $b \approx 4$ as the condition for a single power-law decay in optical contemporaneous to a plateau-to-normal decay transition in X-rays.

A few other bursts show the similar chromatic behavior as in GRB 050319 (e.g., Panaitescu et al. 2006; Oates et al. 2011). However, as pointed by Panaitescu et al. (2006), there is no obvious reason that the evolution of $\eps_e$ and $\eps_B$ has to satisfy such relation as in Eq. (\ref{eq:relation-b-e}). This is the difficulty of this scenario. Alternatively, the chromaticity in these bursts suggests that the X-ray and optical afterglow emission may arise from different origins, for instance, from a two-component outflow in which a low $E_{\rm iso}$, high $\G_j$ component decelerates very early and produces the optical emission with a single power law decay (the LF of this jet component has dropped significantly from its initial $\G_j$, which explains its dominance in the long-wavelength emission), and a high $E_{\rm iso}$, low $\G_j$ component decelerates later at $t_b$ and is responsible for the X-rays.

%%%%%%%%%%%%%%%%%%%%%%%%%%%%%%%%%%%Begin Tab. 2%%%%%%%%%%%%%%%%%%%%%%%%%%%%%%%%%%%%%%%%%%%%%
%\begin{sidewaystable*}
\begin{table*}
\begin{center}
\small
\footnotesize
\caption{The sub-sample that is consistent with our coasting-in-wind model and their model parameter constraints.}      \label{tab:cons}
\vspace{0.6cm}
\begin{tabular}{lrrrrcrcrr}
\hline
 GRB & $t_1$ & $t_b$ & $t_2$ & $F_{\nu_X}(t_b)$ & $z$ & $(\eps_e^{\frac{15}{11}} \eps_B^{\frac{5}{11}})_{\max}$ & $\G_{0, \max}$ & $A_{*, \min}$ & $E_{\rm iso}$ \\
     & ($10^2$ s) & ($10^3$ s) & ($10^4$ s) & ($10^{-7}$ Jy) & & & ($\eps_{e,-1}^{-\frac{21}{88}} \eps_{B,-2}^{\frac{15}{88}}$) & ($10^{-2} \eps_{e,-1}^{-\frac{7}{22}} \eps_{B,-2}^{-\frac{17}{22}}$) & ($10^{52} \eps_{e,-1}^{-\frac{14}{11}} \eps_{B,-2}^{-\frac{1}{11}}$ erg) \\
\hline \hline
 050319  &  61.1  &  11.2  &   8.5  &   9.3  &  3.24  & 109.4  &  49.5  &   3.1  &   7.6 \\
050416A  &   2.5  &   1.7  &  26.2  &  23.8  &  0.65  &   3.5  &  45.1  &   7.7  &   5.1 \\
050713B  &   7.9  &  10.8  &  47.9  &  30.3  &  $\cdots$  &   8.3  &  49.9  &   8.9  &  30.1 \\
 050726  &   4.2  &   1.2  &   1.7  & 104.0  &  $\cdots$  &   3.1  &  80.7  &   2.6  &   6.7 \\
 050801  &   0.7  &   0.3  &   4.6  &  69.9  &  $\cdots$  &   0.7  &  74.0  &   2.0  &   0.8 \\
 050822  &  64.1  &  67.0  &  52.3  &   2.8  &  $\cdots$  & 168.5  &  31.3  &   9.0  &  29.7 \\
051016B  &  47.8  &  66.4  &  15.0  &   1.1  &  0.94  & 188.5  &  26.1  &   8.3  &  20.2 \\
 060109  &   7.4  &   4.9  &   4.8  &  24.3  &  $\cdots$  &   9.2  &  56.7  &   3.8  &   9.7 \\
060204B  &  40.6  &   5.5  &   9.9  &  26.8  &  $\cdots$  &  47.8  &  54.9  &   4.8  &  12.3 \\
 060210  &  39.0  &  24.2  &  86.2  &  18.3  &  3.91  &  47.8  &  48.7  &   7.1  &  30.2 \\
 060306  &   2.5  &   4.7  &  12.4  &  22.8  &  $\cdots$  &   3.2  &  52.9  &   4.6  &   8.6 \\
060428A  &   2.3  &  11.0  &  27.1  &  27.7  &  $\cdots$  &   2.5  &  50.9  &   7.6  &  28.5 \\
 060507  &  30.0  &   7.0  &   8.6  &   8.7  &  $\cdots$  &  57.6  &  45.5  &   3.8  &   5.8 \\
 060526  &  10.9  &  10.0  &  32.3  &   6.1  &  3.21  &  23.9  &  42.4  &   3.9  &   4.6 \\
 060604  &  35.2  &  11.4  &  40.4  &   6.2  &  2.68  &  75.5  &  40.4  &   4.9  &   6.2 \\
 060707  &  53.2  &  22.2  &  81.4  &   1.9  &  3.43  & 186.0  &  32.4  &   4.5  &   3.8 \\
 060708  &  38.1  &   6.7  &  43.9  &   8.4  &  $\cdots$  &  74.6  &  40.9  &   5.6  &   5.3 \\
 060714  &   3.2  &   3.7  &  33.2  &  29.8  &  2.71  &   3.7  &  55.3  &   4.8  &   6.8 \\
 060719  &   2.8  &   9.6  &  18.2  &   9.2  &  $\cdots$  &   5.1  &  43.4  &   5.1  &   8.9 \\
 060729  &   4.2  &  73.0  & 222.1  &  23.6  &  0.54  &   4.1  &  34.8  &  42.1  & 447.9 \\
 060804  &   1.8  &   0.9  &  12.2  & 133.7  &  $\cdots$  &   1.2  &  75.2  &   4.2  &   5.8 \\
060805A  &   2.3  &   1.3  &   7.6  &   5.9  &  $\cdots$  &   6.1  &  45.0  &   2.1  &   0.6 \\
 060813  &   0.9  &   1.8  &   7.4  & 225.1  &  $\cdots$  &   0.4  &  83.0  &   5.1  &  22.0 \\
 060814  &   5.7  &  17.4  &  39.9  &  20.7  &  $\cdots$  &   6.8  &  46.6  &   9.0  &  37.6 \\
 060912  &   4.2  &   1.1  &   8.7  &  31.6  &  0.94  &   5.3  &  53.4  &   4.6  &   3.4 \\
%061202  &   9.3  &  41.6  &  35.7  &  33.0  &  $\cdots$  &   8.3  &  49.6  &  12.5  & 160.6 \\
%Exclude this strong softening burst from the consistent sub-sample.
 070129  &  13.2  &  20.1  &  54.6  &   6.9  &  $\cdots$  &  25.5  &  37.8  &   7.9  &  16.5 \\
\hline
\end{tabular}
\end{center}
%\end{sidewaystable*}
\tablecomments{The subscript ``max'' means the upper limit and ``min'' the lower limit. In deriving those constraints, we assumed a common value set $p= 2.4$ and $k= 2$. Observed red shifts are taken from Liang et al. (2007); for those with unknown red shift, $z= 2$ is assumed. A $H_0$ = 71, $\Omega_{\Lambda}$ = 0.73, $\Omega_M$= 0.27 universe is assumed.}
\end{table*}
%%%%%%%%%%%%%%%%%%%%%%%%%%%%%%%%%%%End Tab. 2%%%%%%%%%%%%%%%%%%%%%%%%%%%%%%%%%%%%%%%%%%%%%

%%%%%%%%%%%%%%%%%%%%%%%%%%%%%%%%%%%%%%%%%%%%%%%%%%%%%%%%%%%%%%%%%%%%%%%%%%%%
\section{Constraints on parameters of the coasting-in-wind model}   \label{sec:Constraints}

In the previous section we examined a large sample of X-ray plateaus and found the coasting-in-wind model with $\max(\nu_m, \nu_c)<\nu_X$ to be consistent with the majority of bursts.  While this model is consistent with the data for all the sample in the plateau phase, it can accommodate only 55\% of sample data in the post-plateau phase. In this section we aim to put constraints on the model parameters based on the consistent sub-sample.

For this model to work, the deceleration time $t_{\rm dec}$ must equal $t_b$; $t_{\rm dec}$ is given by setting $\G(t_{\rm dec}) = \G_0$ in Eq. (\ref{eq:Eiso}). Following from this, the blast wave isotropic energy can be inferred. The synchrotron spectral regime $\max(\nu_m, \nu_c) < \nu_X$ must be justified for both phase 1 and 2. Since with time $\nu_m$ decreases and $\nu_c$ increases monotonically during both phases (Eqs. \ref{eq:scaling-pha-1} - \ref{eq:scaling-pha-2}), the spectral constraints are: $\nu_m(t_1) < \nu_X$ and $\nu_c(t_2) < \nu_X$, where $t_1$ is the observed time of the earliest plateau data point and $t_2$ is that of the latest post-plateau decay data point. The last constraint is that the predicted flux density level has to match the observed one. Without losing generality, we choose to calculate the flux density at $t_b$. Those constraints are summarized as follows.
\begin{itemize}
 \item $t_{\rm dec} = t_b \sim 10^4$ s.

 \item $\nu_m (t = t_1 \sim 500 \,{\rm s}) \leq \nu_X$.

 \item $\nu_c (t = t_2 \sim 10^5 \,{\rm s}) \leq \nu_X$.

 \item $F_{\nu_X} (t = t_b \sim 10^4 \,{\rm s}) \sim 1 ~\mu$Jy.
\end{itemize}

The formulae to calculate the model predictions are
\beq        \label{eq:cons-eiso-gen}
E_{\rm iso}= \frac{4.5 \times 6^{3-k}}{17-4k} \times 10^{52-10k} ~ A \G_0^{8-2k} \left(\frac{t_b}{1+z}\right)^{3-k} ~{\rm erg}.
\eeq
\beq
\nu_m(t_1)= \begin{cases}
   3\times10^{24-5k} \times 6^{-k/2} f_p^2 \eps_e^2 \eps_B^{1/2} \G_0^{4-k} A^{1/2} t_1^{-k/2}\\
   ~~~~~~ \times (1+z)^{k/2-1} ~{\rm Hz}, ~~~~~ {\rm for} ~~ 2 < p, \\
   6.5\times10^{21}\\
   ~~~~ \times (4.8\times 10^{2-5k}\times 6^{-k/2} f_p^2 \eps_e^2 \eps_B^{1/2} A^{1/2})^{1/(p-1)} \\
   ~~~~ \times \G_0^{\frac{p+2-k}{p-1}} t_1^{-\frac{k}{2(p-1)}} (1+z)^{\frac{k}{2(p-1)}-1} ~{\rm Hz},\\
   ~~~~~~~~~~~~~~~~~~~~~~~~~~~~~~~~ {\rm for} ~~ 1 < p < 2, \hfill
  \end{cases}
\eeq
\beq
\begin{split}
\nu_c(t_2) = 0.65\times [6(4-k)]^{3k/2} \times 10^{15k-10} \eps_B^{-3/2} A^{-3/2} \G_0^{3k-4} \\
 \times \left(\frac{t_2}{1+z}\right)^{3k/2-2} (1+z)^{-1} \left(\frac{t_2}{t_b}\right)^{-\frac{(3-k)(3k-4)}{2(4-k)}} ~{\rm Hz}.
\end{split}
\eeq
\beq        \label{eq:cons-fnux-gen}
F_{\nu_X}(t_b)= \begin{cases}
    \frac{ 0.72\times 10^3 \times(3.1\times10^3)^{p-2} \times (6\times10^{10})^{-\frac{(p+2)}{4} k} }{(3-k) (1+z)^{\frac{p}{2}-1} D_{28}^2} \\
    ~~~~ \times f_p^{p-1} \eps_e^{p-1} \eps_B^{\frac{p-2}{4}} A^{\frac{p+2}{4}} \G_0^{(p+2)(2-k/2)} \\
    ~~~~~~~~ \times \left(\frac{t_b}{1+z}\right)^{2-\frac{(p+2)}{4} k}  ~{\rm Jy}, ~~~ {\rm for} ~~ 2 < p, \\
    \frac{0.84\times10^3 \times (1.64\times10^2)^{p-2} \times (6\times10^{10})^{-k}}{(3-k) (1+z)^{\frac{p}{2}-1} D_{28}^2} \\
    ~~~~ \times f_p \eps_e A \G_0^{7+\frac{p}{2}-2k} \left(\frac{t_b}{1+z}\right)^{2-k} ~{\rm Jy}, \\
    ~~~~~~~~~~~~~~~~~~~~~~~~~~~~~~~~~~ {\rm for} ~~ 1 < p < 2,
\end{cases}
\eeq
where we have kept all model and observable parameters in the expressions. The flux density calculation has included a correction factor $\approx 0.2$ due to both the radial internal structure of the blast wave and the prolate equal-arrival-time surface effects (Granot, Piran \& Sari 1999).

In principle, one can calculate these predictions and work out constraints using all the available parameters [$p$, $k$, $F_{\nu_X}(t_b)$, etc.]. However, the appearance of $p$ and $k$ in the exponents would then obscure the dependence on basic parameters such as $\eps_e$ and $\eps_B$, and make comparisons very difficult.   We choose instead to adopt
the most common set of values $k=2$ and $p=2.4$ (corresponding to $\beta_{X,1}=1.2$). Table \ref{tab:cons} lists the individual constraining results for the consistent sub-sample using individual values of $t_1$, $t_2$, $t_b$, $F_{\nu_X}(t_b)$ and $z$.

As an example for demonstration, in the following we use all typical observable values, i.e., $p=2.4$, $k=2$, $t_1= 500$ s, $t_b=10^4$ s, $t_2= 10^5$ s, $\nu_X=$ 1 keV, $z=2$, and $F_{\nu_X}(t_b)= 1$ $\mu$Jy, to carry out the constraints. It turns out the results using common values are consistent with those listed in Table \ref{tab:cons} using individual values. Adopting these values, the items to be constrained become
\beq    \label{eq:cons_eiso}
E_{\rm iso}= 0.5\times10^{44} ~ A_* \G_0^4 t_b ~ {\rm erg}.
\eeq
\beq    \label{eq:cons_nu_m}
\nu_m (t_1= 500 {\rm s}) = 5.7\times10^{15} ~ \eps_e^2 \eps_B^{1/2} A_*^{1/2} \G_0^2 ~ {\rm Hz} \leq \nu_X,
\eeq
\beq    \label{eq:cons_nu_c}
\nu_c (t_2= 10^5 {\rm s}) = 1.1\times10^9 ~ \eps_B^{-3/2} A_*^{-3/2} \G_0^2 ~ {\rm Hz} \leq \nu_X,
\eeq
\beq    \label{eq:cons_fnux}
\begin{split}
F_{\nu_X} (t_b= 10^4 {\rm s}) & = 5.9\times10^{-11} ~ \eps_e^{7/5} \eps_B^{1/10} A_*^{11/10} \G_0^{22/5} ~ {\rm Jy} \\
 & =  1 ~ \mu{\rm Jy},
\end{split}
\eeq
Eq. (\ref{eq:cons_fnux}) gives
\beq    \label{eq:A_*}
A_*= 7.0\times10^3~\eps_e^{-14/11} \eps_B^{-1/11} \G_0^{-4}.
\eeq
Plugging the above into Eq. (\ref{eq:cons_nu_m}), we have
\beq
\eps_e^{15/11} \eps_B^{5/11} \leq 0.9.
\eeq
Various data modeling work and numerical experiments gave various values of $\eps_e$ and $\eps_B$. A fairly conservative range is $\eps_e \sim 0.01 - 0.5$, and $\eps_B \sim 0.0001 - 0.1$, respectively. Thus, the above constraint of $\eps_e$ and $\eps_B$ can be easily met.
Plugging Eq. (\ref{eq:A_*}) into Eq. (\ref{eq:cons_nu_c}) gives an upper limit on $\G_0$:
\beq    \label{eq:cons_G0}
\G_0 \leq 46 ~ \eps_{e,-1}^{-21/88} \eps_{B,-2}^{15/88}.
\eeq
When plugged back into Eq. (\ref{eq:A_*}), the above implies
\beq    \label{eq:cons_A_*}
A_* \geq 4.6\times10^{-2}~\eps_{e,-1}^{-7/22} \eps_{B,-2}^{-17/22}.
\eeq
Finally, plugging Eq. (\ref{eq:A_*}) into Eq. (\ref{eq:cons_eiso}), we have
\beq
E_{\rm iso} = 1.0\times10^{53}~\eps_{e,-1}^{-14/11} \eps_{B,-2}^{-1/11} t_{b,4} ~~ {\rm erg}.
\eeq
For properties of wind from a typical Wolf--Rayet star, the lower limit on $A_*$ in Eq. (\ref{eq:cons_A_*}) is easily met. Therefore, the major model parameter constraints from the above are $\G_0 \lesssim 50$ and $E_{\rm iso} \sim 10^{53}$ erg.

\begin{figure}
\centerline{
\includegraphics[width=9cm, angle=0]{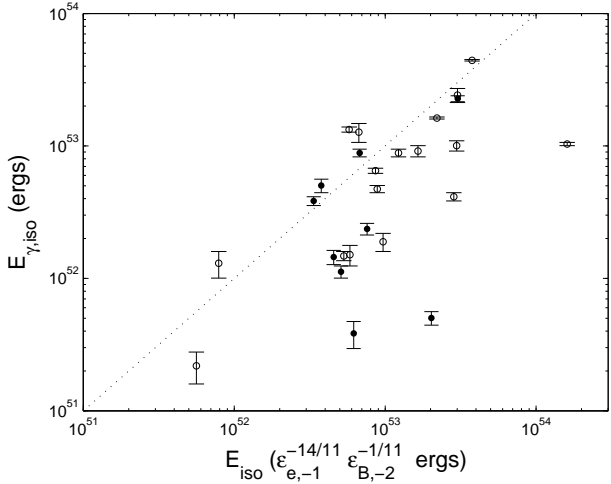}
}
\caption{The total energy release in prompt $\g$-rays vs. the total energy of the ejecta inferred from X-ray data for our consistent sub-sample, both in isotropic equivalent values. \textit{Filled} circles -- bursts with measured $z$. \textit{Open} circles -- bursts with unknown $z$, for which $z= 2$ is assumed. The \textit{dotted} line is $E_{\g,{\rm iso}}=E_{\rm iso}$. }	 \label{fig:Eg-E}
\end{figure}

We plot the individual $E_{\rm iso}$ vs. $E_{\g,{\rm iso}}$ --- the isotropic energy release in prompt $\g$-rays --- for the consistent sub-sample in Figure \ref{fig:Eg-E}. An almost linear correlation between the two can be seen, which is consistent with and likely derives from the observed correlation between the plateau X-ray fluence and the prompt $\g$-ray fluence (Liang et al. 2007). It also shows that $E_{\g,{\rm iso}}$ is comparable to or smaller than $E_{\rm iso}$, which alleviates the troublesome issue one faces in refreshed shock models of extremely high $\g$-ray radiative efficiency $\sim$ 90\% (Zhang et al. 2007) .

For all bursts in the sample, the post-plateau light curve does not show any further steepening break that can be identified as the jet break up to $t_2$. This fact sets a lower limit on the ejecta beaming angle $\theta_j \geq 1/\G(t_2) = \G_0^{-1} (t_2/t_b)^{(3-k)/[2(4-k)]}$. For $k=2$, $t_b \approx 10^4$ s, $t_2 \approx 10^5$ s and using Eq. (\ref{eq:cons_G0}), we obtain $\theta_j \geq 0.04~ \eps_{e,-1}^{21/88} \eps_{B,-2}^{-15/88}$ radians --- not a strong constraint.

%%%%%%%%%%%%%%%%%%%%%%%%%%%%%%%%%%%%%%%%%%%%%%%%%%%%%%%%%%%%%%%%%%%%%%%%%%%%%%%%%%%%%%%%%%%%%
%%%%%%%%%%%%%%%%%%%%%%%%%%%%%%%%%%%%%%%%%%%%%%%%%%%%%%%%%%%%%%%%%%%%%%%%%%%%%%%%%%%%%%%%%%

%%%%%%%%%%%%%%%%%%%%%%%%%%%%%%%%%%%%%%%%%%%%%%%%%%%%%%%%%%%%%%%%%%%%%%%%%%%%%%%%%%%%%%%%%%%%%%
\section{Summary and Discussion}    \label{sec:Discussion}

By analyzing the closure relations for a large sample of X-ray afterglows with plateaus, we find that the plateau feature and post-plateau decay can be explained by the coasting-in-wind model in 55\% of the sample. This simple model is also able to explain the contemporaneously observed optical afterglow emission when optical data are available; however, for a very few bursts for which the end of the plateau is chromatic, additional assumptions (such as evolution of the microphysical parameters $\eps_e$ and $\eps_B$, or a two-component outflow) are needed to explain the lack of a break in the optical light curve.  For the remaining 45\% of the sample, the coasting-in-wind model can still reproduce the plateau, but the post-plateau decay in these bursts is too rapid to be explained in this model; a refreshed shock remains the most capable interpretation.

Given the constraints derived from the consistent sub-sample, the coasting-in-wind model poses two physical challenges which must be addressed.

First, is it reasonable for the wind medium to extend to the large distances $r =  4 \G_0^2 c (t_b t_2)^{1/2}/(1+z) \approx 10^{18}$\,cm, implied by this model?  In fact it is: for a wind speed $10^8 v_{w,8}$\,cm\,s$^{-1}$, the wind ram pressure at this distance is $10^{-8.3} A_* v_{w,8}^2 (10^{18} {\rm cm}/r)^2$\,dyne\,cm$^{-2}$, which exceeds the hydrostatic pressure ($P\simeq G\Sigma^2$) for any column density $\Sigma < 0.3 A_*^{1/2} v_{w,8}$ $( 10^{18}\,{\rm cm}/r )\,{\rm g\,cm}^{-2}$. For the lowest acceptable values of $A_*$ we might therefore expect the wind to have terminated, especially within the high pressures and column densities of a starburst environment.  However, a fiducial wind ($A_*\sim 1$) is sufficient to compete with starburst pressures (note that Dai \& Wu 2003 found $A_* \sim 0.01$ for a burst which lacked an X-ray plateau). In fact, it is quite possible that at these radii the circum-burst medium is a merged wind from an entire star cluster (Chevalier \& Clegg 1985), as massive stars are rarely found alone.   The cluster mass function  (McKee \& Williams 1997) implies that the progenitor star is equally likely to belong to a massive cluster as a small one, and star clusters --- especially massive ones --- are very effective at clearing their immediate molecular environments with winds, radiation pressure, and photoevaporation before any stellar cores collapse (e.g., Krumholz \& Matzner 2009).

One must also consider the expansion of the wind-termination shock as the bubble expands (see Ramirez-Ruiz et al.\ 2005); this is $>10^{18}$\,cm at $10^4$ years if either $n_H<10^2 A_* v_{w,8}^{4/3}$\,cm$^{-3}$ in a uniform medium, or if the bubble expands into a previous stellar wind which is $<10^{2.7} A_* v_{w,8}^{4/3}$ times denser (Koo \& McKee 1992).  However, other cluster stars help to clear the ambient medium and alleviate this constraint. The duration of phase 2 is therefore not a strong constraint on the model, except in cases where there is independent evidence that the wind is weak and the ambient pressure and density are very high.

The second and more puzzling challenge involves the arrangement of ejecta from the central engine.   Our upper limit $\Gamma_0 \lesssim 50$ is well below the lower limit of $10^{2-3}$ derived from the prompt $\g$-rays using the pair opacity constraint (Lithwick \& Sari 2001) for a few {\it Fermi} Large Area Telescope (LAT) bursts, e.g., $\G \gtrsim$ 600 -- 900 for 080916C (Abdo et al. 2009a), $\G \gtrsim 10^3$ for short-hard burst 090510 (Ackermann et al. 2010), $\G \gtrsim 10^3$ for 090902B (Abdo et al. 2009b), and $\G \approx$ 200 -- 700 for 090926A (Ackermann et al. 2011). However, the above constraints assume that both the MeV and GeV emission are produced in the same region; considering two-zone models for these two spectral components reduces the lower limit on $\G$ by a factor of $\sim$ 5 (Zhao, Li \& Bai 2011; Zou, Fan \& Piran 2011; Hasco\"{e}t et al. 2011).

The discrepancy on the $\G$ constraint between the above individual results and ours may be resolved by two means: (1) a possible selection bias. Three of the above four {\it Fermi}-LAT bursts do not have early ($t< 5\times10^4$ s) X-ray or optical data to tell whether a plateau feature is present. The last one, GRB 090510, has X-ray and optical coverage from $t \sim 100$ s and both X-ray and optical light curves show a shallow-to-steep transition at $t \sim 10^3$ s which, however, is found to be consistent with a jet break (Kumar \& Barniol Duran 2009). These four bursts are all among those most energetic ones ($E_{\g,{\rm iso}} \sim 10^{54-55}$ ergs, even the short-hard GRB 090510 has $E_{\g,{\rm iso}}= 10^{53}$ ergs). It is quite plausible that these most energetic bursts have much higher outflow LFs, so that their plateau features are too short ($t_b \propto E_{\rm iso} \G_0^{-4}$; see Eq. \ref{eq:cons-eiso-gen}) to be caught or identified. Future accumulation of Fermi-LAT bursts with early X-ray / optical coverage could either support or disprove this bias effect. (2) The above constraints on $\G$ for each burst are derived from individual pulse(s) during the prompt phase, while our work is for the afterglow phase. It is very natural to have an ejecta bulk LF in the afterglow phase that is much lower than that of an `individual emitter' during the prompt phase. This can be understood in the frames of two major GRB prompt phase models. In the conventional internal shock model (e.g., Paczynski \& Xu 1994; Rees \& M\'{e}sz\'{a}ros 1994), discrete shells with large LF variation among them collide with each other, with each collision corresponding to an individual prompt pulse. Approximately after the prompt phase, all shells merge together and external shocks develop. It is possible that the prompt emission is produced only from the high LF portion of the outflow, while the outflow total energy is dominated by the lower LF portion so that external shock possesses a lower LF. In another category of models that involve `individual emitters' due to, e.g., magnetic turbulence inside a relativistic bulk flow (Lyutikov \& Blandford 2003; Narayan \& Kumar 2009), the prompt-emission-derived LF limit or value is actually for the bulk flow LF \textit{times} the `individual emitters' LF. After the `individual emitters' energy is released in form of $\g$-rays, the external shock LF is just that of the initial bulk flow.

Alternatively the bulk LF may have been reduced by baryon pollution during time between the prompt and afterglow phases.   Such contamination cannot come from the wind, for this would imply that the fast ejecta decelerate early and skip the observed plateau.   Possible stellar sources include: (1) the blowout of an inflated cocoon within the stellar envelope, which may have mixed with an unknown amount of stellar envelope material, and (2) the cap of stellar envelope which is unable to escape the advancing jet head, and is cast forward by the process of runaway shock acceleration (Matzner 2003).   Of these, it is unclear how the former would pose an obstacle to the high-$\G$ jet material after it has escaped the star.

The latter source, a trapped portion of the outer envelope, is worthy of closer inspection.  To evaluate this, we examine the approximate conditions for trapping of matter in front of the jet head (eqs.\ [26] and [27] of Matzner 2003, subject to his eq.\ [6]\footnote{Note the typo (inverted expression) in the middle expression of equation (6) in Matzner (2003).}).   When trapping occurs while the head's motion is nonrelativistic, the result is simple: the jet traps material once it is within $\sim 2.5\theta_j R_*$ of the stellar surface, where $R_*$ is the star's radius; in a polytropic region with index $n$, the trapped mass scales as $\theta_j^{3+n}$ (a very strong function of $\theta_j$).  If instead the head is relativistic, the dependence weakens because a narrower, faster jet traps matter from deeper in the envelope; however it also depends on the intensity as well as the opening angle of the jet.  For an example, consider the outer envelope profile for the SN 1998bw progenitor by Woosley, Eastman, \& Schmidt (1999) and discussed by Matzner (2003, \S\,5.1.1).  In this compact, luminous, helium-stripped Wolf--Rayet star the {\em isotropic} rest energy of trapped material is
\beq
\begin{split}
\max\left[1.8 \left(\frac{\theta}{3^\circ}\right)^5, 1.0 \left(\frac{\theta}{3^\circ}\right)^{2.5} \left( \frac{L_{j,{\rm iso}}} {10^{52}\,{\rm erg\,s}^{-1}}\right)^{0.63} \right]\\
\times 10^{49} ~{\rm erg}.
\end{split}
\eeq
Although this material is swept into an accelerating forward shock (Matzner 2003), a negligible fraction attains LFs higher than $\G_j$.   Moreover, although it is small, the rest energy is well above the critical value $\G_j^{-2} E_{{\rm iso}}$ required to decelerate the ejecta.  Because it is about $10^{-2} (10^{52} \,{\rm erg}/E_{\rm iso})$ of the jet energy (for the fiducial case cited), the final LF after the interaction would be $\sim 10^2$, but strongly dependent on $\theta_j$; our inferred $\Gamma_j \sim 40$ is entirely plausible as a final value.   We refer the reader to the discussion by Thompson (2006), who considers a similar scenario.

Whatever the origin of the lower-$\G_j$ matter, it is clear that it inherits effectively all of the GRB kinetic energy from the outflow powering the prompt phase. First, Figure \ref{fig:Eg-E} shows that the afterglow energy budget is compatible with what is expected from the prompt phase, given reasonable radiative efficiencies.  Second, if any significant fraction of the energy had proceeded beyond the coasting shell at higher LF, it would have decelerated early and caused a noticeable departure from the plateau phase at early times.  Indeed, since in the deceleration phase, when $\max(\nu_m,\nu_c) < \nu_X$ (from Eqs. \ref{eq:cons-eiso-gen} and \ref{eq:cons-fnux-gen}),
\beq
E_{\rm iso} \propto F_{\nu_X}(t_b)^{4/(p+2)} t_b^{(3p-2)/(2+p)},
\eeq
we infer that the persistence of the plateau from times $t_1$ to $t_b$ rules out any early injection of fast ejecta with more than a fraction $(t_1/t_b)^{(3p-2)/(2+p)} \sim 0.03$ of the total energy.

For 45\% of the our sample, the post-plateau decay cannot be explained by phase 2 of the coasting-in-wind model. If we invoke the refreshed shock model for these afterglows, the inferred parameter suggests that the total outflow kinetic energy is dominated by either the late ejecta (this issue might be alleviated for a lepton-dominated late ejecta) or by the slower, massive ejecta. This poses a great theoretical challenge to how the central engine works, and in fact is also the major motivation for us to investigate the coasting-in-wind model which does not face this serious constraint. The standard rate history of the accretion due to the fallback of supernova-shocked stellar material gives a fairly steep slope: $\dot{M} \propto t^{-5/3}$ (Michel 1988; Chevalier 1989; MacFadyen, Woosley \& Heger 2001), which is unable to provide conditions for a refreshed shock. It was recently proposed that an extended central engine activity could be realized due to the continuing accretion of the entire progenitor stellar envelope (Kumar, Narayan \& Johnson 2008a, b) or the slow self-adjustment of a transient debris torus formed at the stellar radius ($\sim 10^{11}$ cm; Cannizzo \& Gehrels 2009). For the second variety of the refreshed shock scenario where a late engine activity is not needed, the slower, massive ejecta that catches up with the decelerating blastwave might actually be the broken-out cocoon that was produced when the GRB jet penetrating through the star (Ramirez-Ruiz, Celotti \& Rees 2002; Matzner 2003).

\vspace{0.8cm}
Note added in proof: an independent investigation of the same physical model, with application to two bursts, has been posted in a preprint by Lei et al. (2011).

\section*{}
We thank the referee for useful comments that helped to improve the paper and Chris Thompson for stimulating suggestions. R.-F.S. thanks Rodolfo Barniol Duran, Enrico Ramirez-Ruiz and Eleonora Troja for useful comments about the manuscript. The research of R.-F.S. and C.D.M. is supported by an NSERC Discovery grant.

\end{document}